\documentclass[12pt]{article}
\usepackage{graphicx}
\begin{document}
\def\la{\langle}
\def\ra{\rangle}
\newcommand{\smfrac}[2]{\mbox{$\frac{#1}{#2}$}}

\begin{center}
\subsection*{
The magnetic reorientation transition in thin ferromagnetic films
treated by many-body Green's function theory}

P. Fr\"obrich$^+$, and P.J. Kuntz

Hahn-Meitner-Institut Berlin, Glienicker Stra{\ss}e 100, D-14109 Berlin,
Germany,\\
$^+$also: Institut f\"ur Theoretische Physik, Freie Universit\"at Berlin\\
Arnimallee 14, D-14195 Berlin, Germany\\

e-mail: froebrich@hmi.de or kuntz@hmi.de
\end{center}

\begin{abstract}
This contribution describes the  reorientation of the magnetization
of thin ferromagnetic Heisenberg films as function of the
temperature and/or an external field. Working in a rotating frame
allows an exact treatment of the single-ion anisotropy when going to higher
order Green's functions. Terms due to the exchange interaction are treated by
a generalized Tyablicov (RPA) decoupling.
\end{abstract}.

\subsection*{1. Introduction}
New experimental results concerning the properties of thin magnetic films have
provided an incentive for theoretical investigations. This contribution deals
with the reorientation of the magnetization of thin ferromagnetic Heisenberg
films as function of the temperature or an external magnetic field. It is of
particular importance to take into account collective excitations (magnons),
which influence the magnetic properties of films more strongly than those of
bulk magnets. Many-body Green's function theory (GFT) is an appropriate tool to
achieve this.
In previous work, we investigated the reorientation transition by applying
GFT \cite{FJK00,FJKE00}, where a Tyablikov
(RPA)-decoupling
of the exchange interaction terms  and a Anderson-Callen decoupling \cite{AC64}
of the single-ion anisotropy terms in the equation of motion for the
lowest-order GF was used. By comparing with Quantum Monte Carlo (QMC) results
\cite{HFKTJ02}
it was shown that this leads to rather good results for small anisotropies
($K_{2}\leq 0.1 J$) when the magnetic field is in the direction of the
anisotropy but the approximation is not as good when the field is applied
perpendicular to the anisotropy. A considerable simplification and an
improvement of the results concerning the reorientation
is reported in Ref.
\cite{SKN05}, where the Anderson-Callen decoupling is made in a frame which
is rotated with respect to the original one and in which the magnetization is
in the direction of the new {\em z}-axis.
The reorientation angle is determined from the condition
that the magnetization commutes  with the Hamiltonian in
the rotated frame. In this connection see also Ref.\cite{PPS05},
who also apply the approximate Anderson-Callen decoupling in a rotated frame.
We realized in Ref. \cite{FKS02}, generalizing findings of Ref.\cite{Dev71},
that when introducing higher-order GF's a decoupling of
the anisotropy terms is not necessary. By taking
advantage
of relations between products of spin operators \cite{JA00}, this leads to an
automatic
closure of the hierarchy of equations of motion with respect to the anisotropy
terms. This procedure gives improved results (as compared to the approximate
Anderson-Callen treatment) when applying
a field in the direction of the anisotropy, which now can be large
\cite{FKS02}, but there are difficulties
(there were zeroes in the equation of motion matrix requiring a special
treatment \cite{FK03,FK05}, and numerical instabilities occurred) when trying
to calculate
the reorientation induced by a  field perpendicular to the anisotropy.
The {\em new} result of the present paper is that these difficulties can
be overcome when working in a rotated frame, in analogy to Ref.\cite{SKN05}. In
this way, we are able to describe the field-induced spin
reorientation transition for spin $S\geq 1$ with an \underline{exact} treatment
of the
single-ion anisotropy. The exchange interaction terms are still decoupled with
generalized RPA.

\subsection*{2. The Green's function formalism}
We investigate a spin Hamiltonian consisting
of an isotropic Heisenberg exchange interaction between nearest
neighbour lattice sites with strength $J_{kl}$, a second-order
single-ion lattice anisotropy with strength $K_{2,k}$
vertical to the film ($x-y$)-plane   and an external magnetic field
${\bf{B}}=(B_0^x,0,B_0^z)$:
\begin{equation}
{\cal
H}=-\frac{1}{2}\sum_{<kl>}J_{kl}\Big(S_k^-S_l^++S_k^zS_l^z\Big)
 -\sum_kK_{2,k}\Big(S_k^z\Big)^2
-\sum_k\Big(B_0^xS_k^x+B_0^zS_k^z\Big).
\label{1}
\end{equation}
In this paper, we restrict ourselves to this Hamiltonian; we dealt with the
dipole-dipole
interaction in Ref.\cite{FJKE00} and with exchange anisotropies in
Ref.\cite{FK03a}.

Owing to the external $B_0^x$-field, the magnetization vector, initially in the
z-direction, will rotate by an angle $\theta$ in the $x-z$-plane, pointing now
in the $z'$-direction of a new frame $(x',y',z')$. As in Ref.\cite{SKN05}, we
shall do the calculations in the primed system, in which the magnetization
vector has the components $(0, 0, \la S^z\ '\ra)$. The transformation between
the frames is
\begin{equation}
\left(
        \begin{array}{c}
        \la S^x\ra \\
        \la S^y\ra \\
        \la S^z\ra
\end{array}\right ) =  \left( \begin{array}{ccc}
                             \cos\theta & 0 & \sin\theta \\
                              0         & 1 & 0          \\
                             -\sin\theta& 0 & \cos\theta
\end{array} \right)
\left( \begin{array}{c}
       \la S^x\ '\ra \\
       \la S^y\ '\ra \\
       \la S^z\ '\ra
\end{array} \right).
\label{2}
\end{equation}
Because $\la S^x\ '\ra=\la S^y\ '\ra=0$ in the rotated frame, one needs only to
calculate $\la S^z\ '\ra$ in order to find the components of the magnetization
in the original frame provided the angle $\theta$ is known.

After transforming the Hamiltonian (\ref{1}) to the primed system the following
Green's functions are needed
\begin{eqnarray}
G_{ij}^{+,-}&=&\la\la S_i^+\ ';S_j^-\ '\ra\ra\ , \nonumber\\
G_{ij}^{(z)^n+,-}&=&\la\la (S_i^z\ ')^{n-1}(2S^z_i\ '-1)S_i^+\ ';S_j^-\ '\ra\ra
\ . \label{3}
\end{eqnarray}
The single-ion anisotropy is active for spins $S\geq 1$, and one needs the
first Green's function plus those for $n=1, 2, 3, ...$  to
treat films with $S=1, 3/2, 2, ...$.

In establishing the equations of motion, the exchange interaction terms are
treated by a generalized Tyablikov (RPA)-decoupling, in which we do not break
products of spin operators with equal indices
\begin{equation}
\la\la (S_i^z\ ')^nS_k^+\ ';S_j^-\ '\ra\ra \simeq \la (S_i^z\ ')^n\ra \la
\la S_k^+\ ';S^-_j\ '\ra\ra
+ \la S_k^+\ '\ra \la\la (S_i^z\ ')^n; S_j^-\ '
\ra\ra \ .
\label{4}
\end{equation}
Note that in the rotated system, $\la S^+_k\ '\ra=0$; i.e. the second term
vanishes.
After applying the decoupling procedure and performing a Fourier transform to
momentum space one obtains the following set of equations of motion.
%(For convenience we leave out the prime symbol at the spin operators)
%
\begin{eqnarray}
& &\omega G^{+,-}= 2\la S^z\ '\ra +\la S^z\ '\ra J(q-\gamma_{\bf
k})G^{+,-}\nonumber\\
& &\ \ +(B_0^x\sin\theta+B_0^z\cos\theta)G^{+,-}
+K_2(1-\smfrac{3}{2}\sin^2\theta)G^{z+,-},\nonumber\\
& &\omega G^{z+,-}=(6\la (S^z\ ')^2\ra -2S(S+1))
 -\smfrac{1}{2}J\gamma_{\bf k}\Big(6\la (S^z\ ')^2\ra
-2S(S+1)\Big)G^{+,-}\nonumber\\
& &\ \ +Jq\la S^z\ '\ra G^{z+,-}
+(B_0^x\sin\theta+B_0^z\cos\theta)G^{z+,-}
\nonumber\\
& &\ \ +K_2(1-\smfrac{3}{2}\sin^2\theta)\Big(2G^{(z)^2+,-}-G^{z+,-}\Big)\
, \nonumber\\
& &\omega G^{(z)^2+,-}= 8\la (S^z\ ')^3\ra+3\la(S^z\ ')^2\ra
 -(4S(S+1)-1)\la S^z\ '\ra -S(S+1)\nonumber\\
& &\ \ +J\gamma_{\bf k}\Big(\smfrac{1}{2}S(S+1)+(2S(S+1)-1)\la S^z\
'\ra -\smfrac{3}{2}\la (S^z\ ')^2\ra -4\la(S^z\ ')^3\Big)G^{+,-}
\nonumber\\ & &\ +Jq\la S^z\ '\ra G^{(z)^2+,-}
 +(B_0^x\sin\theta+B_0^z\cos\theta)G^{(z)^2+,-}\nonumber\\
& &\ \ +K_2(1-\smfrac{3}{2}\sin^2\theta)\Big(2G^{(z)^3+,-}-G^{(z)^2+,-}\Big)\ ;
\nonumber\\
& &\omega G^{(z)^3+,-}=10\la (S^z\ ')^4\ra +8\la (S^z\ ')^3\ra
 -(6S(S+1)-5)\la(S^z\ ')^2\ra\nonumber\\
& &\ \ -(4S(S+1)-1)\la S^z\ '\ra -S(S+1)\nonumber\\
& &\ \ +J\gamma_{\bf k}\Big(\smfrac{1}{2}S(S+1)+(2S(S+1)-\smfrac{1}{2})\la
S^z\ '\ra
 +(3S(S+1)-\smfrac{5}{2})\la (S^z\ ')^2\ra\nonumber\\
& &\ \  -4\la (S^z\ ')^3\ra -5\la
(S^z\ ')^4\ra\Big)G^{+,-}\nonumber\\
& &\ \ +Jq\la S^z\ '\ra G^{(z)^3+,-}
+(B_0^x\sin\theta+B_0^z\cos\theta)G^{(z)^3+,-}\nonumber\\
& &\ \ +K_2(1-\smfrac{3}{2}\sin^2\theta)\Big(2G^{(z)^4+,-}-G^{(z)^3+,-}\Big)\ .
\label{5}
\end{eqnarray}
For a square lattice with lattice contant $a=1$, the quantity
$\gamma_{\bf k}=2(\cos{k_x}+\cos{k_y})$, and $q=4$, the number of nearest
neighbours.

As Ref.\cite{SKN05} we neglect all
GF's not containing an equal number of $S^-\ '$
and $S^+\ '$ operators.
%  because the corresponding correlations are
%nondiagonal with respect to the magnetic quantum number.

One observes that in  the equations (\ref{5}) the anisotropy terms do not yet
lead to a closed system. This is, however, achieved by
using formulas derived in Ref.\cite{JA00}, which reduce products of spin
operators by one order (!).
One obtains
\begin{eqnarray}
\rm{for}\ \  S=1:\ \ \ &
&G^{(z)^2+,-}=\smfrac{1}{2}(G^{z+,-}+G^{+,-})\ ,\nonumber\\
\rm{for}\ \  S=3/2:\ \ \ &
&G^{(z)^3+,-}=G^{(z)^2+,-}+\smfrac{3}{4}G^{z+,-}\ ,\nonumber\\
\rm{for}\ \  S=2:\ \ \ &
&G^{(z)^4+,-}=\smfrac{3}{2}G^{(z)^3+,-}+\smfrac{7}{4}G^{(z)^2+,-}\nonumber\\
& &-\smfrac{9}{8}G^{z+,-}-\smfrac{9}{8}G^{+,-}\ .
\label{6}
\end{eqnarray}
When inserting these relations into the system of equations (\ref{5}) one
sees that the resulting system of equations is closed.

The equations of motion can be written in compact matrix notation
\begin{equation}
(\omega{\bf 1}-{\bf \Gamma}){\bf G}={\bf A}.
\label{7}
\end{equation}
The quantities ${\bf \Gamma,\ G,\ {\rm and}\  A}$ can be read off from equation
(\ref{5}), where the {\em non-symmetric} matrix ${\bf \Gamma}$ is a
$(2\times 2),\ (3\times 3),\ (4\times 4)$ -matrix for spins $S=1,\ 3/2,\ 2$,
respectively.
The desired correlation vector corresponding to the Green's functions
(\ref{3}),
\begin{equation}
{\bf C}=\left( \begin{array}{c}
\la S^-\ 'S^+\ '\ra \\ \la S^-\ 'S^{(z')^{n-1}}(2S^z\ '-1)S^+\ '\ra
\end{array}
\right),
\label{8}
\end{equation}
is obtained via the spectral theorem. Using the eigenvector method of
Ref.\cite{FJKE00}, one has for the components of the correlation vector
${\bf C}$ after a Fourier transform to configuration space
\begin{eqnarray}
\label{9}
C_i&=&\int d{\bf k}C_i({\bf k})
=\frac{1}{\pi^2}\int_0^\pi dk_x\int_0^\pi
dk_y\sum_{j,k,l=1}^{2S}R_{ij}\epsilon_{jk}L_{kl}A_l\\
& & (i=1,2,..,2S),\nonumber
\end{eqnarray}
where the integration is over the first Brillouin zone, and $\bf R(L)$ are
matrices whose columns (rows) consist of the right (left) eigenvectors of the
matrix ${\bf \Gamma}$, and
$\epsilon_{jk}=\delta_{jk}/(e^{\beta \omega_j}-1)$ is a diagonal matrix,
in which $\omega_j$ are the eigenvalues ($j=1,..,2S$) of the ${\bf
\Gamma}$-matrix.

It remains to derive an equation which determines the rotation angle.
This is again done  by using the approximation that the
commutator of the magnetization with the Hamiltonian vanishes
in the rotated system.
 %
%\begin{equation}
%\frac{d}{dt}{S^z\ '}=\sum_i
%\label{10}
%\end{equation}
%
This implies that  the following Green's function is zero
\begin{equation}
\la\la [S_i^z\ ',H']; S_j^-\ ' \ra\ra=0.
\label{11}
\end{equation}
Evaluating the commutator, one obtains
\begin{equation}
(B_0^x\cos\theta-B_0^z\sin\theta)G_{ij}^{+-}-K_2\sin\theta\cos\theta
G_{ij}^{z+,-}=0.
\label{12}
\end{equation}
After applying the spectral theorem to this equation one obtains for the
diagonal
correlations the relation which determines the reorientation angle angle:
\begin{equation}
(B_0^x\cos\theta-B_0^z\sin\theta)C^{-+}-K_2\sin\theta\cos\theta
C^{-z+}=0\ .
\label{13}
\end{equation}
This is the generalization of the angle condition of Refs. \cite{SKN05,PPS05},
when treating the single-ion anisotropy exactly instead of applying the
Anderson-Callen decoupling.
Equation (\ref{13}) together with the set of integral equations (\ref{9}) have
to be
solved self-consistently in order to obtain the magnetization $\la S^z\ '\ra$
and
its moments in the rotated system together with the reorientation angle
$\theta$. Applying the relations (\ref{2}) yields the
components of the magnetizations in the original system in which the magnetic
reorientation is measured.

\subsection*{3. Results}
We solve the equations of the last section with the curve following method
described in detail in an appendix of Ref. \cite{FKS02}.
In Ref. \cite{SKN05}, where the Anderson-Callen decoupling was used in the
rotated frame, only cases with weak single-ion anisotropy were considered.
An example was the case of spin $S=2$ and $K_2=0.01J$, which is a rather small
anisotropy as appearing for 3d transition metals. A very good agreement with
QMC calculations of Ref. \cite{HFKTJ02} was obtained. Because the anisotropy is
small the result with the present theory is practically indistinguishable,
whereas the results of Ref.\cite{FJK00} are rather bad for the reorientation,
because the decoupling was not done in the rotated frame.
%\newpage
\begin{figure}[htp]
\begin{center}
\protect
\includegraphics*[bb=80 90 540 700,
angle=-90,clip=true,width=12cm]{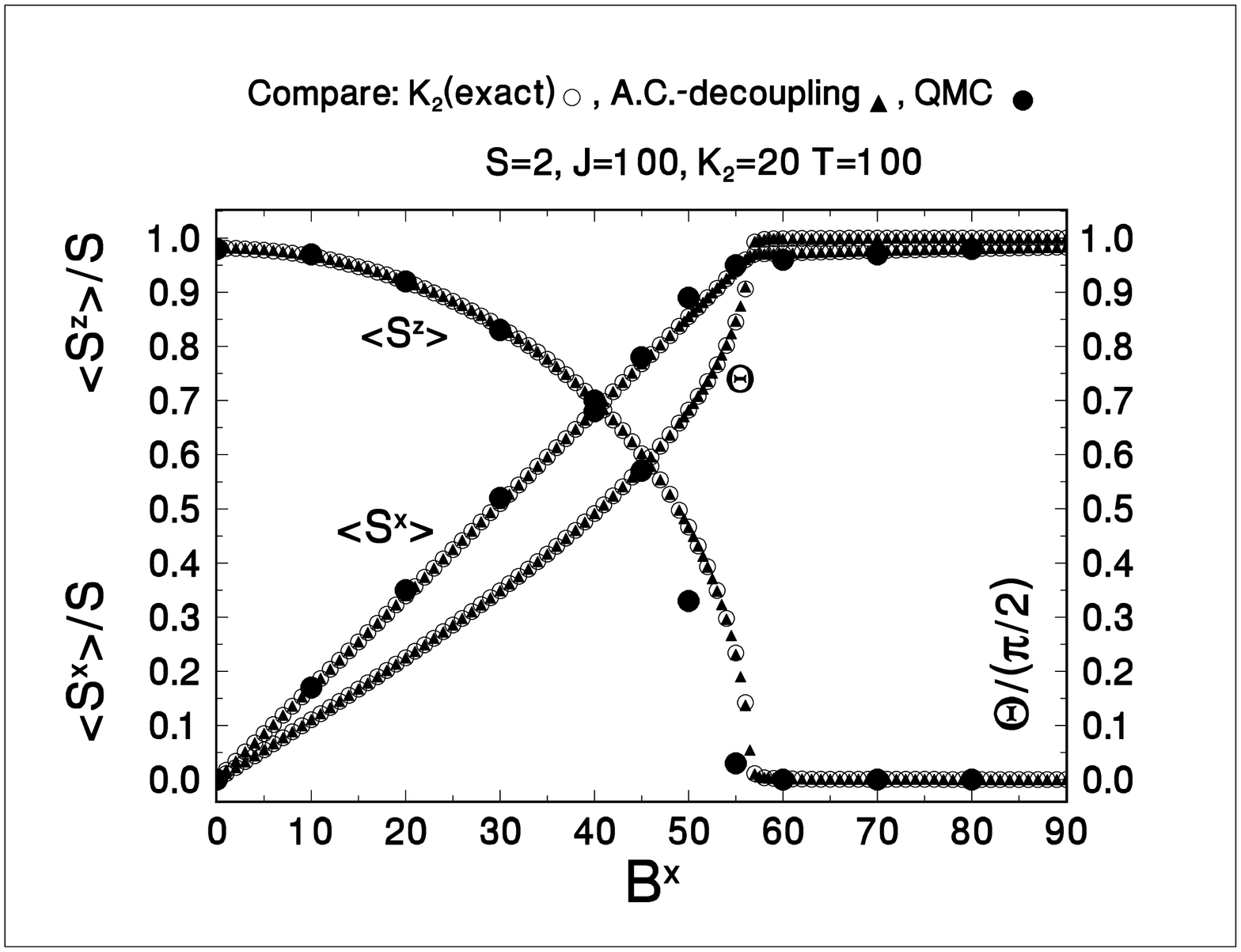}
\protect
\caption{Normalized magnetizations $\la S^z\ra/S$ and $\la S^x\ra/S$ and the
reorientation
angle $\Theta$ for a spin S=2 Heisenberg monolayer as function of the
external field are shown: QMC \cite{HFKTJ02}(solid circles), Anderson-Callen
decoupling \cite{SKN05}(triangles), present theory (open circles).}
\vspace{-0.5cm}
\end{center}
\end{figure}
For anisotropies in the rare earth region, which can be of the order of the
exchange interaction, the approximate theory of Ref. \cite{SKN05} should break
down, and one expects
the present theory to be superior. Surprisingly, the Anderson-Callen decoupling
in the rotated frame still yields excellent results, when compared with the
present theory and QMC results from Ref. \cite{HFKTJ02} for anisotropies up to
$K_2\leq 0.2J$. This can be seen
in Fig. 1 for the magnetic reorientation  induced by the transversal
$B^x$-field for the case of $K_2=0.2J$ and $T=100$. The result of both Green's
function theories are nearly identical to each other and deviate only slightly
from the exact( within the statistical error) Quantum Monte Carlo results.
 %
%\vspace{-2cm}

%
\begin{figure}[htp]
\begin{center}
\protect
\includegraphics*[bb=80 100 530 680,
angle=-90,clip=true,width=12cm]{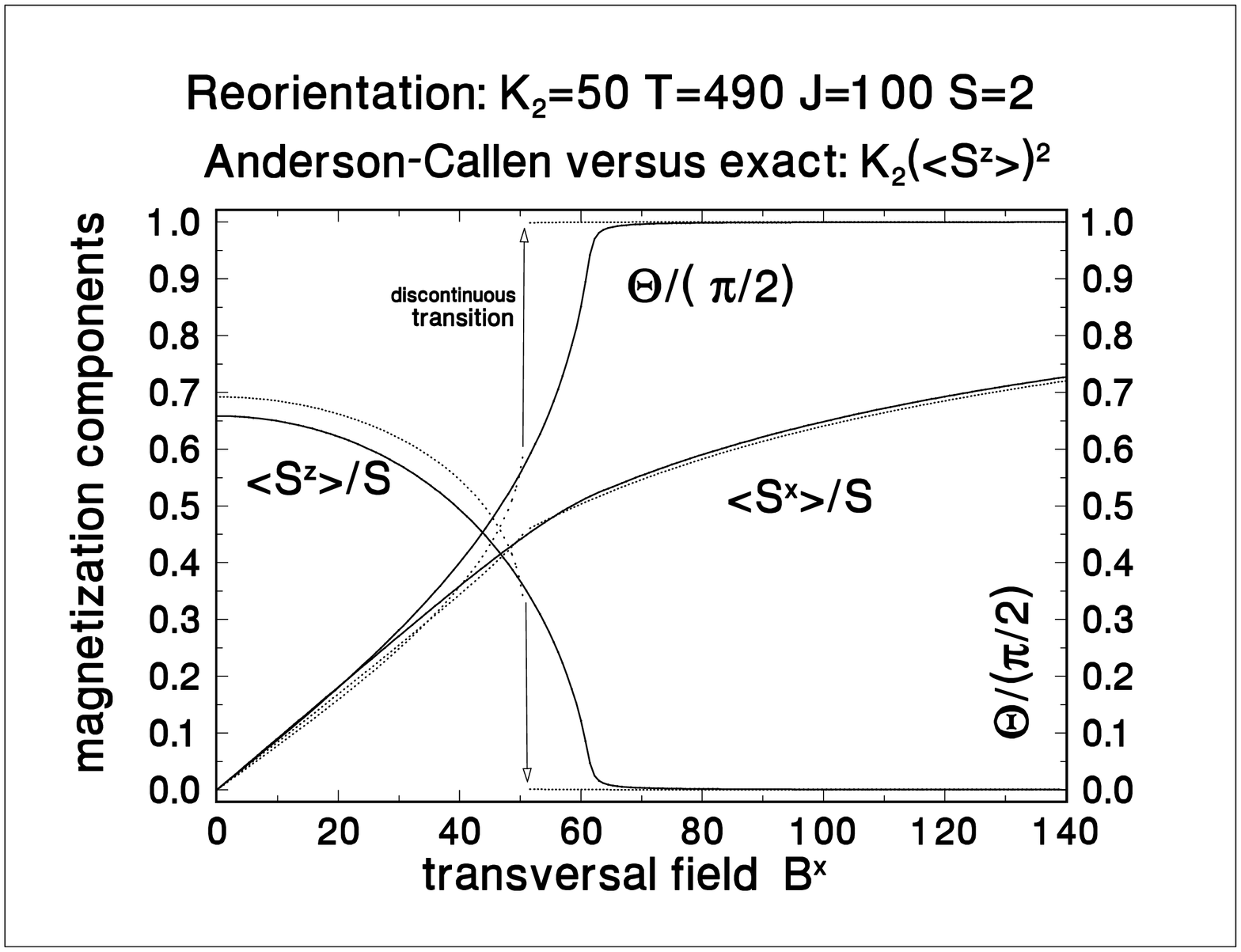}
\protect
\caption{Normalized magnetizations $\la S^z\ra/S$ and $\la S^x\ra/S$ and the
reorientation
angle $\Theta/(\pi/2)$ for a spin S=2 Heisenberg monolayer as function of a
transversal field $B^x$ are shown:  Anderson-Callen decoupling
\cite{SKN05}(dotted lines) and the present theory (solid lines)
for $ K_2=0.5J\ (T/J=4.9)$.}
\end{center}
\end{figure}
%
%\vspace{-1cm}
%

The reason that the magnetization curves as function of the external field of
both theories
are very close for T=100 is that with the present choice of the parameters
the magnetizations at this temperature are still very close to each other. Both
theories deviate from each other
with increasing temperature, until close to the reorientation temperature the
deviation is
maximal. Therefore one should observe deviations between the theories in this
temperature range. The differences should also increase with increasing
anisotropy strength $K_2$. Therefore we have calculated the field-induced
reorientation  for  a temperature somewhat below the reorientation temperature
and for a large anisotropy
$ K_2=0.5J\ (T/J=4.9),$ for a
Heisenberg monolayer with spin $S=2$.
The result is shown in Fig. 2.
In this
case the Anderson-Callen (A.C.)
decoupling along
the lines of \cite{SKN05}
leads to a discontinuous transition from a certain
angle ($\theta/(\pi/2)\approx 0.6$) to full reorientation
($\theta/(\pi/2)=1$). whereas the reorientation
transition is continuous when the single-ion anisotropy is treated exactly.
 Such discrete transitions are also reported in Ref.
\cite{PPS05} in a treatment which is very similar
to that of Ref. \cite{SKN05}. The reason why
this is not observed in the last reference is that there
only very small anisotropies were considered. We attribute the discontinuous
transition to the approximate Anderson-Callen decoupling, which is not
justified when going to large anisotropies.
The difference between the corresponding reorientation fields, $B_{R}$,
increases with increasing
anisotropy. For the present case it is:
$B_R^{\rm present}-B_R^{\rm A.C.}\simeq 11\  (\rm{for}\ K_2=0.5J)$.

Unfortunately, we cannot say how accurate the present model is because we have
no QMC calculations available for large anisotropies. The most uncontrolled
approximation
consists in the generalised RPA decoupling of the exchange interaction terms
of the higher-order GF, eqn.(\ref{4}) .
Previous calculations \cite{EFJK99} have shown (by comparing with QMC) that RPA
is a good
approximation for a Heisenberg model (no anisotropy) with a field perpendicular
to the
film plane. For improving the the present approach for a field in
transversal direction one may try to extend the present theory by
applying the procedure of \cite{JIRK04} which goes beyond the RPA with respect
to the exchange interaction terms.
It is possible by applying formulas from Ref.\cite{JA00} to treat also the
fourth-order anisotropy term, $-\sum_iK_{4,i}(S_i^z)^4$, exactly. Work in this
direction is in progress, as well as the generalization to multilayers.
We mention that the reorientation of the magnetization of a ferromagnetic film
induced by the coupling to an antiferromagnetic film can also be treated with
the GF formalism \cite{FK05a}.

\end{document}